\documentclass[fleqn,12pt,twoside]{article}
\usepackage{espcrc1}

\usepackage{graphicx}
\usepackage[figuresright]{rotating}

\newcommand{\AmS}{{\protect\the\textfont2
  A\kern-.1667em\lower.5ex\hbox{M}\kern-.125emS}}

\title{Neutrons and particularities of nucleogenesis 
at early stage of the Universe evolution}

\author{N. Takibayev\address{Center of Basic and Ecological Research, Almaty, Kazakhstan}}
       
\begin{document}

\maketitle

\begin{abstract}

At the early stage of the Universe evolution $t \geq 1$ min., when photons and neutrinos are no longer able to prevent nucleosynthesis, the key role is given to neutron component of matter.
Neutron component creates a certain variety of the lightest chemical 
elements and disappears leaving a sufficient 
portion of Helium isotopes and some tiny portions of Li and Be isotopes. 
Free neutrons disappear in energetic flame of these reactions not using even 
a quarter of their own lifetime. Further the Universe matter has been evolving smoothly without free neutrons.

The following physical principle is taken as a basis: interactions of the most fast component in speed are considered as main and taken into account first. Substance component that meets this requirement provides a thermodynamic equilibrium in the system. It is obvious that fast component of the substance is the neutrons. Other components cause slow change as compared with actions of the main one.

Thermodynamic description of nucleus matter is carried out in the same way as it is made in the well-known problem of ``ionization equilibrium'' of atomic plasma.
\end{abstract}

\section{THERMODYNAMIC EQUILIBRIUM IN FIRST MINUTES}

At this period of the early Universe the lightest nuclei generation is especially effective at temperatures $Ò \sim 0.3 \div 0.1 MeV$  
 \cite{Fowl72,Kolb90,Bedn02,Strg03}.
 
Considering thermodynamic description of the substance the following physical principle is foundation for interactions of the fastest component  are to be considered in the first place. This is due to the fact that substance component provides a thermodynamic equilibrium in the system.  
Speeds of neutron reactions with nuclei are rather exceeding those of protons. This is the reason why the neutrons with all-penetrating character cause thermodynamic equilibrium in the matter faster than others.
The key advantage is independence of thermodynamic values from the details of interactions between particles of the matter. These values are defined by integral characteristics (average energy or temperature, pressure, etc.) and statistical weights of initial and final states \cite{Land76}.

The well-known method of atomic plasma ``ionization equilibrium'' description comes to the following \cite{Land76}:
 
Gas  is considered to consist of neutral atoms at low temperatures. When temperature increasing atoms get ionized:

\begin{equation}
A_0\rightarrow A_1 + e ,\ \   A_1\rightarrow A_2 + e ,\ \  A_2\rightarrow A_3 + e ,\ \  \cdots
\end{equation}
where $A_0$  - neutral atom, $A_1$ - once ionized, $A_2$ - twice ionized and so on.  
As it is applied to these reactions the law of Acting Mass gives a set of well-known Saha's equations  

\begin{equation}
\frac{c_{n-1}}{c_n c} = P K_n(T) ,\ \  n = 1, 2, \ldots 
\end{equation}
where $c_0$ means neutral atom concentration, $c_1, c_2, ...$  - different ion concentrations, but $c$  is electron concentration. The equation $c = c_1 + 2c_2 + 3c_3 + \cdots$  reflects electrical neutrality of gas in general.

As it is known, equilibrium constants can be described as follows:

\begin{equation}
K_n = \frac{g_{n-1}}{2 g_n}\left(\frac{2\pi}{m_e}\right)^{3/2} T^{5/2}
\exp(I_n/T)
\end{equation}
where $g$ is statistical weight of atom (or ion), $m_e$ - mass of electron, but $I_n = \epsilon_{0,n} - \epsilon_{0, n-1}$ - energy of $n$ - ionization. This set of equations defines a concentration of different ions and gives their functional dependence on temperature change.

In our issue the substance components' correlation changes not due to atom ionization but because of the nuclear reactions with neutrons, and the temperature falls not rises.

It is more convenient to divide nuclei into isotope groups: $H$ group -  $(p, d, t)$, $He$ group - $(^3He, ^4He)$, $Li$ group - $(^{6 - 9}Li)$, etc.
 
Neutrons play the role of electrons. 

At the final stage all neutrons are caught by nuclei. Ionization reaction chain can be recorded as follows for each isotope-group:

\begin{equation}
A\rightarrow (A - 1) + n ,\   (A - 1)\rightarrow (A - 2) + n , \ 
(A - 2)\rightarrow (A - 3) + n ,\  \cdots
\end{equation}

It is obvious that the most accurate solution of the problem requires taking into account the nuclear reactions with proton participation and nuclei-nuclei reactions.

However, at the first step one can omit reactions between charged particles because of the stated role of neutrons - as main component, which quickly creates thermodynamic equilibrium in the system. Role of superfluous protons can be considered as a substance - solvent. 
   
In case of chemical equilibrium, solvent is passive in reactions between chemical reagents. Under these circumstances another peculiarity of the problem appears dealing with an issue concluded in each isotope group, which can be considered separately. It means that within one isotope group neutrons can provide a thermodynamic equilibrium very fast. 

At the same time the reactions between nuclei of different isotope groups are ``slow'' processes connected with overcoming the Coulomb barrier.

\section{NUCLEAR ``IONIZATION EQUILIBRIUM'': $H$ GROUP}

Reaction chain in hydrogen isotope group is: $p + n \rightarrow d + \gamma  , \   d + n \rightarrow t + \gamma $.
 
If it is taken other way (i.e. from right to left), it gives an equation for equilibrium constants:

\begin{equation}
\frac{c_t}{c_d c_n} = PK_1(T) ,\ \ \ \  \frac{2c_d}{c_n(c_n - c_d)} = PK_2(T) 
\end{equation}

Here $c_t$  is triton concentration, $c_d$ means that of deutrons, $c_n$ - neutrons. 
Equilibrium constants are as follows:

\begin{equation}
PK_1 = \frac{1}{3}M(T)\exp(I_1/T) ,\ \ \ I_1 = \epsilon_t - \epsilon_d \approx
6,25 MeV ,
\end{equation}
and
\begin{equation}
PK_2 = \frac{3}{4}M(T)\exp(I_2/T) ,\ \ \  I_2 = \epsilon_d \approx 2,23 MeV ,
\end{equation}
where for the adiabatic process: 

\begin{equation}
M(T)\approx C_M T^{-0,01} \approx 10^{-6} .
\end{equation}

An initial quantity $c_n \approx 0,13$ \cite{Bedn02}. Then, coordinated values of concentrations can be gained according to Eqs. (5).

\begin{table}[htb]
\caption{Concentrations of $H$-group isotopes at different temperature.}
\label{table:1}
\newcommand{\m}{\hphantom{$-$}}
\newcommand{\cc}[1]{\multicolumn{1}{c}{#1}}
\renewcommand{\tabcolsep}{2pc} 
\renewcommand{\arraystretch}{1.2} 
\begin{tabular}{@{}lllll}
\hline
$T$ (MeV)     & \cc{$0.3$} & \cc{$0.25$} & \cc{$0.2$} & \cc{$0.1$} \\
\hline
$c_n$         & \m0.13 & \m0.05 & $0.7\cdot10^{-2}$ & $\sim10^{-8}$ \\
$c_d$ 				& $0.63\cdot10^{-5}$ & $0.3\cdot10^{-4}$ & $1.3\cdot10^{-6}$ & $\sim10^{-13}$ \\
$2c_t$     & $2.7\cdot10^{-4}$ & \m0.08 & \m0.12  & \m0.13  \\
\hline
\end{tabular}\\[2pt]

\end{table}
 
It should be noted that $c_t$ rapidly grows from negligible quantity to a high value comparable with initial concentration of free neutrons. 
Then, in the nearest future triton will decay and reform into nuclei $^3He$ . It may be the reason of rather high abundance of $^3He$  at the Universe. 
It is important that at $T < 0,25 MeV$ neutron component is rapidly degenerating. 

\section{NUCLEAR ``IONIZATION EQUILIBRIUM'': $He$ AND $Li$ GROUPS}

In $He$ group the key reaction is to be considered: $^3He + n \rightarrow ^4He + \gamma$ , then:

\begin{equation}
\zeta(T) = \frac{c_4}{c^2_3} = P K(T)\approx \frac{C_M}{4}\exp(20.57/T)  
\end{equation}
where $c_{3,4}$  are concentrations of corresponding isotopes, and $T$ in MeV as above. 

It is witnessing that $\zeta$ is dramatically increasing when $T$ is down from $\zeta(0.3) \sim 10^{23}$ \  to \  $\zeta(0.2) \sim 10^{38}$ \  and \ $\zeta(0.1) \sim 10^{84}$, for example.

The chain of transformations is the following: $^6Li + n \rightarrow ^7Li + \gamma$ , $^7Li + n \rightarrow ^8Li + \gamma$	, $^8Li + n \rightarrow ^9Li + \gamma$, results in the set of Eqs. like (5) - (7). 

Initial quantity of $^6Li$ indicated as $c_6$ is taken in accordance with a simple Boltsmann's distribution \cite{Bedn02}.

\begin{table}[htb]
\caption{Concentrations of $Li$-group isotopes at different temperature.}
\label{table:2}
\newcommand{\m}{\hphantom{$-$}}
\newcommand{\cc}[1]{\multicolumn{1}{c}{#1}}
\renewcommand{\tabcolsep}{2pc} 
\renewcommand{\arraystretch}{1.2} 
\begin{tabular}{@{}lllll}
\hline
$T$ (MeV)     & \cc{$0.3$} & \cc{$0.25$} & \cc{$0.2$} & \cc{$0.1$} \\
\hline
$c_6$         & $\sim10^{-9}$ & $\sim10^{-11}$ & $\sim10^{-14}$ & $\sim10^{-28}$ \\
$c_7$ 				& $\sim10^{-12}$ & $\sim10^{-10}$ & $\sim10^{-10}$ & $\sim10^{-11}$ \\
$2 c_8$     & $\sim10^{-21}$ & $\sim10^{-17}$ & $\sim10^{-16}$ & $\sim10^{-15}$ \\
$3 c_9$     & $\sim10^{-27}$ & $\sim10^{-22}$ & $\sim10^{-18}$ & $\sim10^{-10}$ \\
\hline
\end{tabular}\\[2pt]

\end{table}

\section{CONCLUSION}

\begin{enumerate}
	\item Free neutrons are very quickly caught by the nuclei. This process leads to a drastic increase of senior isotope concentrations in every isotope group.

Free neutrons disappear in energetic flame of nuclear reactions not using even a quarter of their own lifetime. 

Further, the Universe matter has been evolving smoothly without free neutrons.
At that time the more possible reactions occur: $p + t \rightarrow ^4He + \gamma$ and $p + d \rightarrow ^3He + \gamma$, etc.
	\item In $H$ isotope group a number of tritons abruptly increases. A great number of tritons may be a reason of $^3He$ abundance in nature at the present.
	\item  In $He$ isotope group $^3He$ becomes $^4He$ immediately. 
	\item  An increase of $Li$ senior isotope's concentration can contribute in solving the issue of $Be$ primordial abundance.
		
\end{enumerate}

\end{document}